\documentclass[iop,apj,numberedappendix]{emulateapj}
\usepackage{amssymb}
\usepackage{amsmath}
\usepackage{graphicx}
\usepackage{xspace}
\usepackage{natbib}
\usepackage{url}
\usepackage{paralist}
 \usepackage{aasmacros}
\usepackage{multirow}

\slugcomment{ }

\shorttitle{The Broad-Line Radio Galaxy 3C\,390.3}
\shortauthors{Lohfink et al.}

\begin{document}

\title{The Corona of the Broad-Line Radio Galaxy 3C\,390.3}

%%%%%%%%%%%%%%AUTHORS%%%%%%%%%%%%%%%%%%%%%%%%%%%%%%%%%%%%%%%%%%%%%

\author{A.M. Lohfink\altaffilmark{1}, P. Ogle\altaffilmark{2}, F. Tombesi\altaffilmark{3,4}, D. Walton\altaffilmark{5}, M. Balokovi\'{c}\altaffilmark{6}, A. Zoghbi\altaffilmark{7}, D.R. Ballantyne\altaffilmark{8}, S.E. Boggs\altaffilmark{9}, F.E. Christensen\altaffilmark{10}, W.W. Craig\altaffilmark{9,11}, A.C. Fabian\altaffilmark{1}, C.J. Hailey\altaffilmark{10}, F.A. Harrison\altaffilmark{6}, A.L. King\altaffilmark{12}, G. Madejski \altaffilmark{12}, G. Matt\altaffilmark{13}, C.S. Reynolds\altaffilmark{3}, D. Stern\altaffilmark{5}, F. Ursini\altaffilmark{13,14}, W.W. Zhang\altaffilmark{4}}
\email{alohfink@ast.cam.ac.uk}
\altaffiltext{1}{Institute of Astronomy, University of Cambridge, Madingley Road, Cambridge CB3 0HA, UK}
\altaffiltext{2}{Infrared Processing \& Analysis Center, California Institute of Technology, Pasadena, CA 91125, USA}
\altaffiltext{3}{Department of Astronomy, University of Maryland College Park, College Park, Maryland 20742, USA}
\altaffiltext{4}{X-ray Astrophysics Laboratory, NASA/Goddard Space Flight Center, Greenbelt, MD, 20771, USA}
\altaffiltext{5}{Jet Propulsion Laboratory, California Institute of Technology, 4800 Oak Grove Drive, Pasadena, CA 91109, USA}
\altaffiltext{6}{Cahill Center for Astronomy and Astrophysics, Caltech, Pasadena, CA 91125, USA}
\altaffiltext{7}{Department of Astronomy, University of Michigan, 1085 South University Avenue, Ann Arbor, MI 48109, USA}
\altaffiltext{8}{Center for Relativistic Astrophysics, School of Physics, Georgia Institute of Technology, 837 State Street, Atlanta, GA 30332-0430, USA}
\altaffiltext{9}{Space Sciences Laboratory, University of California, Berkeley, 7 Gauss Way, Berkeley, CA 94720-7450, USA}
\altaffiltext{10}{Danish Technical University, DK-2800 Lyngby, Denmark}
\altaffiltext{11}{Lawrence Livermore National Laboratory, Livermore, CA, USA}
\altaffiltext{12}{Kavli Institute for Particle Astrophysics and Cosmology, Stanford University, 2575 Sand Hill Road, MS 29, Menlo Park, CA 94025, USA}
\altaffiltext{13}{Dipartimento di Matematica e Fisica, Universit\`{a} degli Studi Roma Tre, via della Vasca Navale 84, I-00146 Roma, Italy}
\altaffiltext{14}{Univ. Grenoble Alpes, IPAG, 38000, Grenoble, France}

%%%%%%%%%%%%%%%%%%%%%%%%%%ABSTRACT%%%%%%%%%%%%%%%%%%%%%%%%%%%%%%%%%%%%%%%%%%%%%

\begin{abstract}
We present the results from a joint \textit{Suzaku}/\textit{NuSTAR} broad-band spectral analysis of 3C\,390.3. The high quality data enables us to clearly separate the primary continuum from the reprocessed components allowing us to detect a high energy spectral cut-off ($E_\text{cut}=117_{-14}^{+18}$\,keV), and to place constraints on the Comptonization parameters of the primary continuum for the first time. The hard over soft compactness is 69$_{-24}^{+124}$ and the optical depth 4.1$_{-3.6}^{+0.5}$, this leads to an electron temperature of $30_{-8}^{+32}$\,keV. Expanding our study of the Comptonization spectrum to the optical/UV by studying the simultaneous \textit{Swift}-UVOT data, we find indications that the compactness of the corona allows only a small fraction of the total UV/optical flux to be Comptonized. Our analysis of the reprocessed emission show that 3C\,390.3 only has a small amount of reflection ($R\sim0.3$), and of that the vast majority is from distant neutral matter. However we also discover a soft X-ray excess in the source, which can be described by a weak ionized reflection component from the inner parts of the accretion disk. In addition to the backscattered emission, we also detect the highly ionized iron emission lines Fe\,XXV and Fe\,XXVI.   
\end{abstract}


%%%%%%%%%%%%%%START PAPER%%%%%%%%%%%%%%%%%%%%%%%%%%%%%%%%%%%%%%%%%%%%

\section{Introduction}\label{intro}

Broad-line radio galaxies are the radio-loud kin of Seyfert 1 galaxies, and as such they offer the possibility of investigating why some AGN show strong radio jets. Understanding the origin of the radio-loudness is important, since the mechanical energy deposited through AGN jet activity is thought to be a primary way that the AGN can impact its host galaxy evolution \citep{McNamara2007,Fabian2012}.
 Recent studies show that the cool gas necessary to re-trigger the AGN activity needed to power the radio-loudness in radio galaxies, which are pre-dominatently hosted by ellipticals, is provided by galaxy interactions such as mergers \citep{Best2012,Tadhunter2014}. However, this still does not address the issue that, while some AGN do launch powerful jets, many do not. It is possible that the magnetic field build-up and black hole spin are decisive factors for this \citep{Blandford1977,McKinney2009,Sikora2013}. X-ray observations can provide both the determination of the black hole spin, as well as the detection of any changes in the inner accretion disk as one might expect them from significant changes to the magnetic flux strength.

Previous works in the X-ray band have indicated that a typical radio galaxy X-ray spectrum possesses a relatively flat photon index, weak cold reflection signatures, and in some cases a weak relativistic reflection component \citep[e.g.,][]{Grandi2002,Ogle2005,Ballantyne2007,Sambruna2009,Walton2012}. The distinctly flatter photon index raises questions as to whether the entire Comptonization spectrum is different in this source class, implying differences in the balance of cooling and heating of the corona. The recent launch of the \textit{NuSTAR} observatory \citep{Harrison2013} provides the opportunity to study the broadband spectra of AGN with unprecedented sensitivity, allowing us to start directly addressing this issue in sources. To yield the best results the target's X-ray spectrum should possess a continuum dominated by Comptonization, requiring only weak or no jet contribution. A recent study of a suitable source, the radio galaxy 3C\,382, with \textit{NuSTAR} confirms the previous results of weak reflection and a flat photon index. Additionally they show for the first time that the high-energy cut-off of the Comptonization spectrum is time-variable in radio galaxies and possibly flux-dependent \citep{Ballantyne2014}.   

This Paper builds on this discovery and the need for a better understanding of Comptonization spectrum in broad-line radio galaxies, focusing on the coronal parameters of the radio galaxy, 3C\,390.3 ($z=0.056$, $\log(M/M_\odot)=9.04\pm0.05$; \citealt{Grier2013}), using a joint soft X-ray (\textit{Suzaku}) and hard X-ray (\textit{NuSTAR}) observation.

Previous work in the X-ray band on 3C\,390.3 has found its timing properties not to differ from those of radio-quiet Seyferts \citep{Gliozzi2009} and there to be no noticeable contribution from jet to the X-ray emission \citep{Sambruna2009}. \citet{Sambruna2009} performed a detailed broad-band spectral analysis including \textit{XMM}, \textit{Suzaku}, and \textit{Swift}-BAT data, determining that two different spectral models can describe the X-ray spectra well. The first model includes a continuum modeled by a cut-off power law ($\Gamma=1.72\pm0.02$, $E_\text{cut}=161^{+75}_{-62}\,\text{keV}$), cold reflection ($R=0.81\pm 0.04$) and a narrow Gaussian line to model the FeXXV line detected in the data. The second model is similar to the first model but models the FeXXV line with a highly ionized, blurred reflector ($\xi\sim2700$). It has been shown by \citet{Tombesi2013} that the modeling of FeXXV and ionized reflection can be ambiguous in broad-line radio galaxies. 

The paper is organized in the following way: We begin by providing an overview of how the data were reduced in \S \ref{datared}. This is followed by a description of the X-ray spectral modeling (\S \ref{spec_modeling}), and a brief study dedicated to the iron K band (\S \ref{fe_modeling}). The X-ray analysis is then supplemented by a study of spectral energy distribution of this AGN (\S \ref{sed_modeling}) and conclude with a discussion of our results (\S \ref{disc}).  

\section{Data Reduction}\label{datared}
%This work describes the analysis of a joint \textit{XMM}/\textit{Suzaku} and \textit{NuSTAR} observations. 
An overview of the observations considered in this work is given in Table~\ref{nu_obs}. 
\begin{table}
\begin{center}
\caption{Overview of observations and their characteristics. The exposures are the net exposures used during the X-ray broad-band spectral modeling given per XIS in the case of \textit{Suzaku} and per FPM for \textit{NuSTAR}.}\label{nu_obs}
\begin{tabular}{|c|c|c|c|}
\hline Observatory & ObsID & Start time & Exposure \\
 \hline NuSTAR & 60001082002 & 2013-05-24 09:16:07 & \multirow{2}{*}{100} \\
NuSTAR & 60001082003 & 2013-05-24 19:51:07 &  \\
Suzaku & 708034010 & 2013-05-24 07:12:24 & 57 \\
Swift & 00080221001 & 2013-05-24 00:41:59 & -- \\
\hline
\end{tabular}
\end{center}
\end{table}

\subsection{NuSTAR}
The raw NuSTAR data were reduced in the standard manner, using the {\tt nupipeline} and {\tt nuproducts} scripts available in the NuSTARDAS software package\footnote{http://heasarc.gsfc.nasa.gov/docs/nustar/analysis/NuSTARDAS\_swguide\_v1.5.pdf}. We used NuSTARDAS version 1.4.1 and CALDB version 20131223\footnote{\textbf{We have verified that the spectra do not change significantly when the newer CALDB version (20150702) is used.}}. We extracted the source spectrum from a circular region with a 70\arcsec radius, centered on the peak of the point-source image. The background spectra for each FPM were extracted from a polygonal region covering the detector which contains the source, avoiding the area 90\arcsec around the peak.
The total exposure for the data taken in the standard (SCIENCE) mode is 72~ks per module; 24~ks from the first and 48~ks from the second pointing. 

We additionally used the data from the SCIENCE\_SC mode, taken at times when the front star tracker is blinded by the bright Earth limb. Because 3C\,390.3 is at high declination, the fraction of data acquired in this mode is significant. At those times the reconstruction of the target image is based on various combinations of the other three star trackers located on the spacecraft bus. As pointing solutions differ between each combination, therefore slightly changing the placement of the target in the focal plane, we isolated GTIs for each one individually and extracted the source and background spectra in the same manner as in the case of SCIENCE mode data. All high-level products and response files are generated self-consistently for the SCIENCE\_SC mode by the {\tt nuproducts} script. Finally, we co-added the spectra from both pointings and both modes for each FPM separately, totaling 100~ks of exposure per module. For the purpose of this work, we ignore the minimal hardening of the source between the first and second pointing and only use the average spectrum.

\subsection{Suzaku}
The 3C\,390.3 \textit{NuSTAR} data were supported by a simultaneous \textit{Suzaku} observation (ObsID:708034010, PI Ogle). We analyzed the XIS data as described in the \textit{Suzaku} ABC guide in similar fashion to \citet{Lohfink2013}. The PIN data were first screened using \texttt{aepipeline} and the spectra then extracted with \texttt{hxdpinxbpi}. For the background the tuned version was used. The source was too faint in hard X-rays to be detectable by the GSO, only 35\,\% of the counts in the 15-35\,keV band are from the source in the PIN spectrum.

\begin{figure}[ht!]
\vspace*{2.5\baselineskip}
\includegraphics[width=\columnwidth]{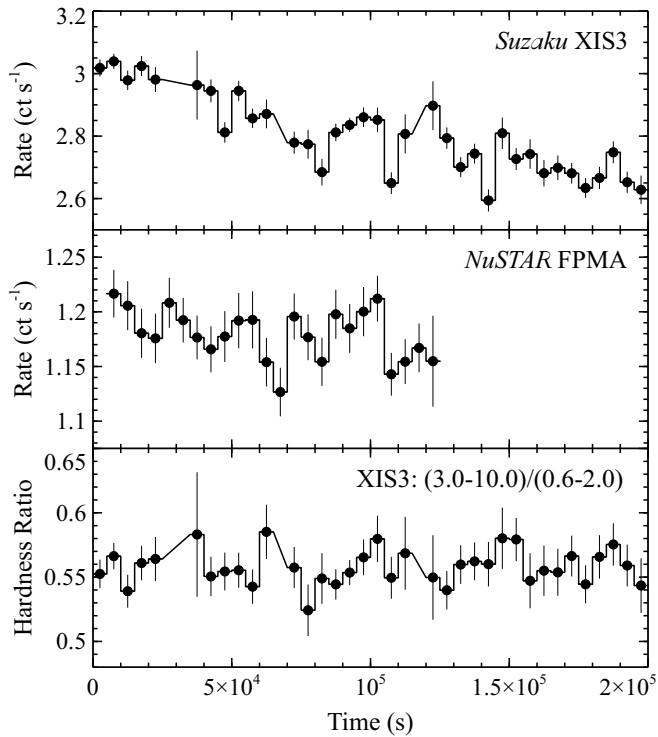}
\caption{\textit{Suzaku}-XIS [top panel] and \textit{NuSTAR} [middle panel] light curves in 5\,ks binning. A slight flux decrease throughout the observations is apparent. The hardness ratio within the \textit{Suzaku}-XIS band [bottom panel] stays constant during this decline.}\label{lightcurve}
\end{figure}

With an elapsed time of 196\,ks, the \textit{Suzaku} pointing is longer than the \textit{NuSTAR} observation (Figure~\ref{lightcurve}), we therefore use spectra reduced with two different good time intervals: one which is simultaneous with \textit{NuSTAR}, and one that includes all times. Throughout the paper the two front-illuminated XIS spectra were combined to one spectrum.

\subsection{Swift}
During the joint \textit{Suzaku} and \textit{NuSTAR} observation, a short snapshot (2.1\,ks) with \textit{Swift} was also taken (ObsID:00080221001). From this observation, we only analyze the UV/optical data from the UVOT instrument as the X-ray data from \textit{Swift}-XRT is affected by pile-up. 3C\,390.3 was observed in three different filters with UVOT: V, U, and W2. The fluxes were extracted from the level II images, using the tool \texttt{uvotsource}. For the source region a circular region around the target's coordinates with 5 arcsec radius was selected, and the background region chosen to be also circular (r=13 arcsec) in a region which is source-free and close to the center of the field-of-view. Galactic reddening was removed assuming an $E(B-V)$ of 0.0616 \citep{Schlegel1998} utilizing a reddening law by \citet{Cardelli1989} updated by \citet{O'Donnell1994}. Other than reddening we also need to account for the host galaxy flux contribution to each of the filter band fluxes. We follow the procedure used in \citet{Lohfink2014}. As the UVOT images are insufficient to perform a decomposition of the AGN and host galaxy contributions, we use the results from HST imaging as a basis. HSTs high quality images allow a detailed 2D modeling of the entire galaxy including the AGN. \citet{Bentz2009} find that the total host galaxy flux of 3C\,390.3 at 5100\,\AA\, is $0.945\times10^{-15}\,\text{ergs}\,\text{s}^{-1}\,\text{cm}^{-2}\,\text{\AA}^{-1}$ and 48\,\% of this flux is from the bulge of the galaxy. We assume for our analysis that our 5 arcsec source region contains the entire bulge and 50\,\% of the remaining galaxy, i.e. about 75\,\% of the total galaxy flux. We use the flux point at 5100\,\AA\, as an anchor to renormalize SED templates by \citet{Kinney1996}. During this work we assume 3C\,390.3 to be an Sa spiral \citep{Bentz2009}. We then convolve the SED templates with the filter functions of our filters and obtain the host galaxy flux values for each of our filters. In general the host galaxy contribution is very small, in the V-band the host galaxy contributes 16\,\% of the observed flux during our observation. For our analysis we then subtract the host galaxy flux in each filter from the de-reddened fluxes.    

\section{Broad-band X-ray Spectral Modeling}\label{spec_modeling}

We first focus on the simultaneous \textit{Suzaku}/\textit{NuSTAR} spectra. The light curve presented in Figure~\ref{lightcurve} shows no strong short-term variability, either in hardness or flux. We therefore only consider the average, simultaneous spectrum at this time. To account for any cross-calibration flux offsets, we always include cross correlation constants in all our fits. The constants are normalized to XISf and in case of PIN the constant is fixed to 1.16. In general, \textit{NuSTAR} and \textit{Suzaku} spectra are known to agree well as a recent study by \citet{Walton2014} also pointed out. For the spectral modeling, the \textit{Suzaku}-XIS and \textit{NuSTAR} spectra are binned to signal-to-noise ratio of 10 to carry equal weight in the fitting. 
Binning to a signal-to-noise of 10 is not possible for the \textit{Suzaku}-PIN data and we therefore require only a signal-to-noise of 5. The energy ranges considered for the modeling are: 0.7-1.7\,keV \& 2.0-10.0\,keV for \textit{Suzaku}-XIS, 15-30\,keV for \textit{Suzaku}-PIN and 3-70\,keV for \textit{NuSTAR}-FPMA/FPMB. 

\begin{figure}[ht!]
\vspace*{2.5\baselineskip}
\includegraphics[width=\columnwidth]{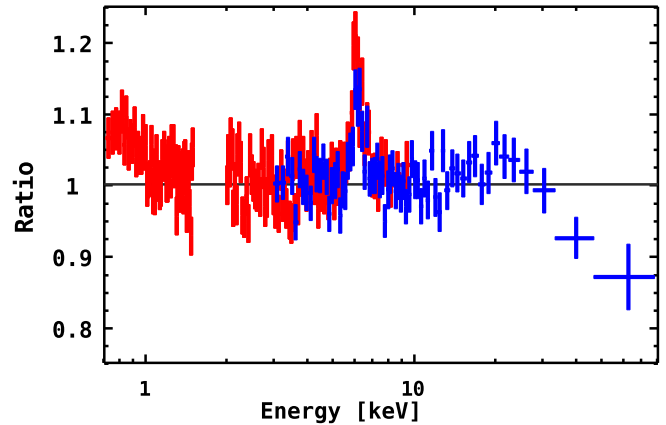}
\caption{Residuals to a simple absorbed power law fitted to the 2-4\,keV and 8-10\,keV band for the front-illuminated \textit{Suzaku}-XIS [red] and \textit{NuSTAR}-FPMA [blue]. The data have been rebinned strongly for plotting.}\label{explore}
\end{figure}

To begin we fit a simple absorbed power law to the 2-4\,keV and 8-10\,keV band. The Galactic absorption is modeled by the \texttt{TBnew} model fixing the $N_\text{H}$ value to the value returned by the $N_\text{H}$ Tool\footnote{https://heasarc.gsfc.nasa.gov/cgi-bin/Tools/w3nh/w3nh.pl} of $4\times10^{20}$\,cm$^{-2}$ \citep{Kalberla2005}. We find a photon index of 1.69. The residuals to the combined front-illuminated \textit{Suzaku}-XIS, \textit{Suzaku}-PIN and \textit{NuSTAR}-FPMA spectra are shown in Figure~\ref{explore}. The residuals indicate the presence of weak reflection in the object, which does appear to be mostly or entirely neutral, as indicated by the iron line energy. A small excess flux at soft X-rays is also apparent in the residuals. 

Based on this initial exploration, we start our more detailed spectral modeling with a model composed of a power law continuum and neutral reflection, both are modeled using the \texttt{pexmon} model and modified by Galactic absorption. We only allow the photon index, and reflection fraction to vary during the modeling. The inclination, iron abundance and high-energy cutoff are initially fixed at these respective values (27\,degrees, 1 solar, 1000\,keV). The inclination of $27\pm2$\,deg was determined from observations of the jet\citep{Dietrich2012}. This simple description is already able to account for most of the residuals seen in Figure~\ref{explore} and results in a $\chi^2$ of 2799.1 for 2585 dof. However at high energies a drop-off is observed and at soft energies some residuals also remain. To account for the high energy drop we allow the high energy cut-off to change and $\chi^2$ improves by 
%2799.1-2530.7=
181.4 for 1 additional free parameter. \textbf{We so initially obtain a best fit high energy cut-off of 110$_{-11}^{+13}$\,keV.} Varying the iron abundance leads to an even better fit ($\Delta \chi^2=13.2$). The iron abundance is found to be slightly super solar (1.6$_{-0.3}^{+0.5}$). Driven by the previous detection of FeXXV we also include a photoionized emitter in the modeling. An XSTAR table is used for the fitting, which is calculated from the model \texttt{photemis}. In our modeling we find that the column density and the normalization of the photoionised emission component are completely degenerate, so we fix the column to $1\times10^{22}\,\text{cm}^{-2}$. The photoionized emission provides a significant improvement to the fit ($\Delta \chi^2=21.5$ for -2 dof) leading to a reduced $\chi^2$ of 1.00. The resulting fit parameters are those typical for a radio galaxy, displaying a low reflection fraction ($R=0.22_{-0.03}^{+0.03}$) and a flat photon index of 1.72$_{-0.01}^{+0.01}$. The photoionized material is found to be highly ionized, i.e. FeXXV and FeXXVI dominate. However none of the already included model components can account for the soft excess seen in the \textit{Suzaku}-XIS spectrum. Guided by previous results we test whether the inclusion of blurred ionized reflection from the central parts of the accretion disk can improve the fit. In the modeling it is represented by the \texttt{relxill} model \citep{Garcia2014} with the reflection fraction set to -1 (to turn of the otherwise included power law). The photon index of the irradiating powerlaw is assumed to be the same than the primary continuum. Further, the emissivity profile is taken to be a simple power law. The inclination of the accretion disk is fixed to the known inclination of 27\,degrees and black hole spin is allowed to vary freely. We find the $\chi^2$ improves to 2555.4 for 2577 dof ($\Delta \chi^2=27.6$ for -4 dof). The inclusion of ionized reflection only alters the parameters slightly, they mostly remain within the errorbars of the previous fit as the ionized reflection only predominantly models the soft excess.  We find that the data are not able to constrain the black hole spin. The emissivity index however is determined to be 2.0$_{-0.0}^{+0.7}$ and the ionization state is found to be moderate with $\log(\xi)=2.0_{-0.2}^{+0.7}$.  

\begin{table*}[ht!]
\begin{center}
\caption{Spectral fit parameters from the modeling of the \textit{Suzaku} and \textit{NuSTAR} data of 3C\,390.3. Column one shows the best fit parameters of the best fit where \texttt{pexmon} provides the continuum, in column two \texttt{compTT} provides the continuum and in column three \texttt{eqpair} provides the continuum.  Parameters marked with an 'f' have been kept fixed at their previous best fit values. Detailed descriptions of the models can be found in the text.}\label{3c390_fit}
\begin{tabular}{c|c|c|c|c}
\hline \hline Cold Reflection & $\Gamma$ & 1.71$_{-0.01}^{+0.01}$ & 1.71f & 1.71f \\
 & $E_\text{cut}$ [keV] & 117$_{-14}^{+18}$ & 117f & 117f \\
 & $R$ &  0.11$_{-0.03}^{+0.04}$ &  0.09$_{-0.03}^{+0.03}$ & 0.30$_{-0.03}^{+0.03}$ \\
 & Fe/Solar & 2.3$_{-0.6}^{+0.8}$ & 4.0$_{-1.1}^{+1.0}$ & 0.8$_{-0.1}^{+0.1}$\\
 & $N_\text{pex}\, [10^{-2}]$ & 1.00$_{-0.03}^{+0.01}$ & 1.00f & 1.00f \\
\hline Photoemission & $N_\text{phot}$ & $>0.004$ & $>0.003$ & $0.006_{-0.004}^{+0.006}$ \\
                & $\log(\xi)$ & 4.2$_{-0.4}^{+0.4}$ & 4.2$_{-0.4}^{+2.5}$ & 4.0$_{-0.2}^{+0.2}$ \\
\hline Ionized Reflection & $N_\text{rel}$ [$10^{-4}$] & 11.9$_{-4.9}^{+3.7}$ & 6.6$_{-2.0}^{+3.5}$ & 22.8$_{-1.5}^{+2.1}$ \\
&$a$ &  uncon. & uncon. & 0.993$_{-0.003}^{+0.003}$ \\
& $q$ & 2.0$^{+0.7}$ & 2.0$^{+0.6}$ & $>9.8$ \\
& $\log(\xi)$ & 2.0$_{-0.2}^{+0.7}$ & 2.5$_{-0.5}^{+0.2}$ & 3.6$_{-0.1}^{+0.1}$ \\
\hline Comptonization & $kT_e$ [keV] & \nodata   & 16$_{-2}^{+4}$ & $30_{-8}^{+32\star}$ \\ 
 & $\ell_h/\ell_s$ & \nodata & \nodata & 69$_{-24}^{+124}$ \\
 & $\tau$ &\nodata & 5.5$_{-0.6}^{+0.4}$ & 4.1$_{-3.6}^{+0.5}$ \\
 & $N_\text{compTT} [10^{-3}]$ & \nodata & 14.1$_{-2.5}^{+1.4}$ & \nodata \\
 & $N_\text{eqpair} $ & \nodata & \nodata & 100$_{-60}^{+53}$ \\
\hline Cross Calibration & $c_\text{xis1}$ &  0.976$_{-0.006}^{+0.005}$ & 0.976f & 0.976f\\
  & $c_\text{pin}$ & 1.16f & 1.16f & 1.16f\\
  & $c_\text{fa}$ & 1.00$_{-0.01}^{+0.01}$ & 1.00f & 1.00f\\
  & $c_\text{fb}$ & 1.05$_{-0.01}^{+0.01}$ & 1.05f & 1.05f\\
\hline \hline  & $\chi^2$ & 2555.4 & 2560.4 & 2564.3\\
  & dof & 2577 & 2580 & 2580\\
\end{tabular}
\tablenotemark{$^\star$ While the coronal temperature is not a free parameter in the \texttt{eqpair} model it can be displayed during the fitting. This is the best fit value.}
\end{center}
\end{table*}

\begin{figure*}[ht!]
\begin{minipage}{0.49\textwidth}
\includegraphics[width=\columnwidth]{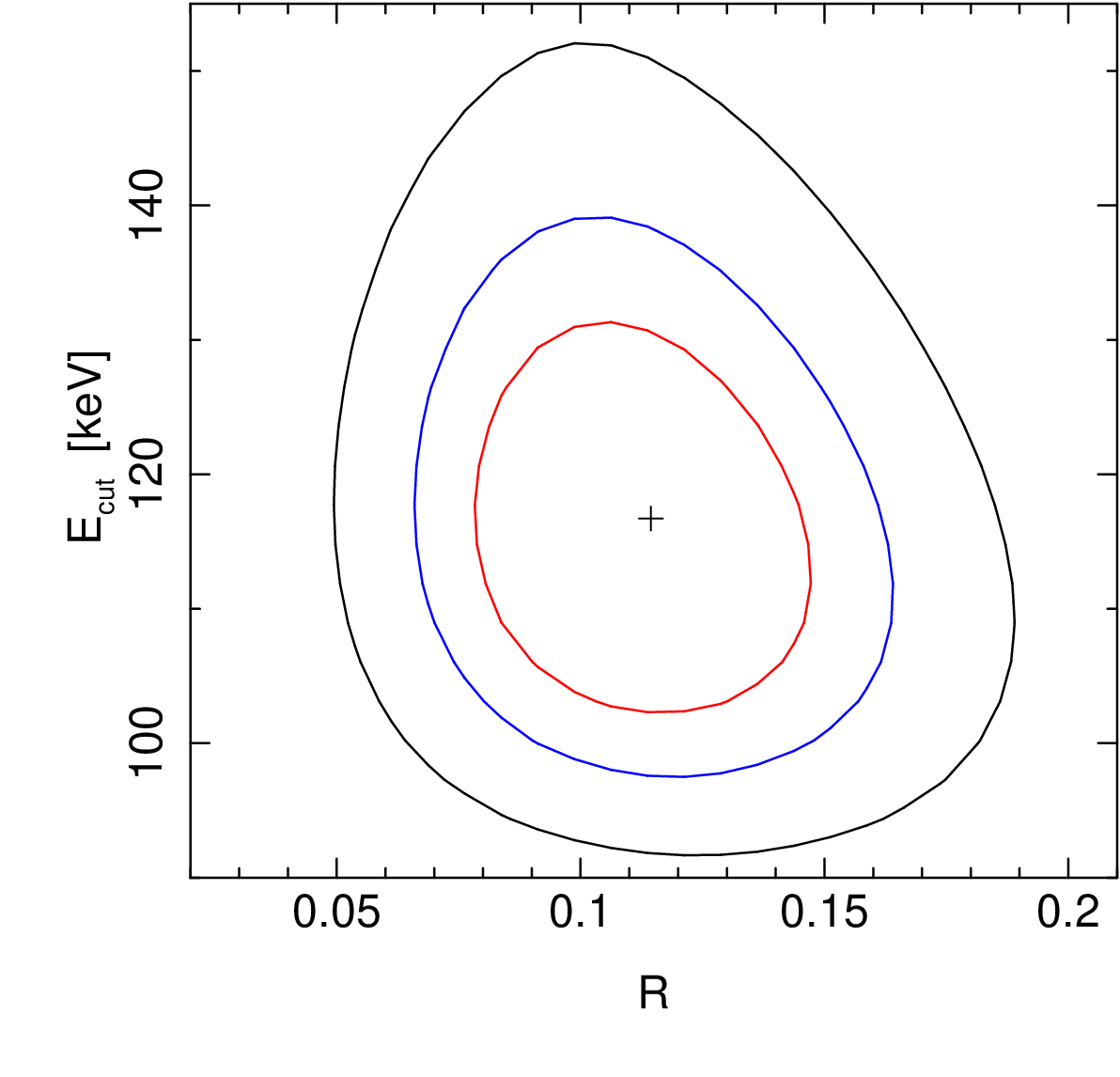}
\caption{Confidence contour of the reflection fraction $R$ and the high energy cut-off energy $E_\text{cut}$.}\label{r_ecut}
\end{minipage}
\hfill
\begin{minipage}{0.49\textwidth}
\includegraphics[width=\columnwidth]{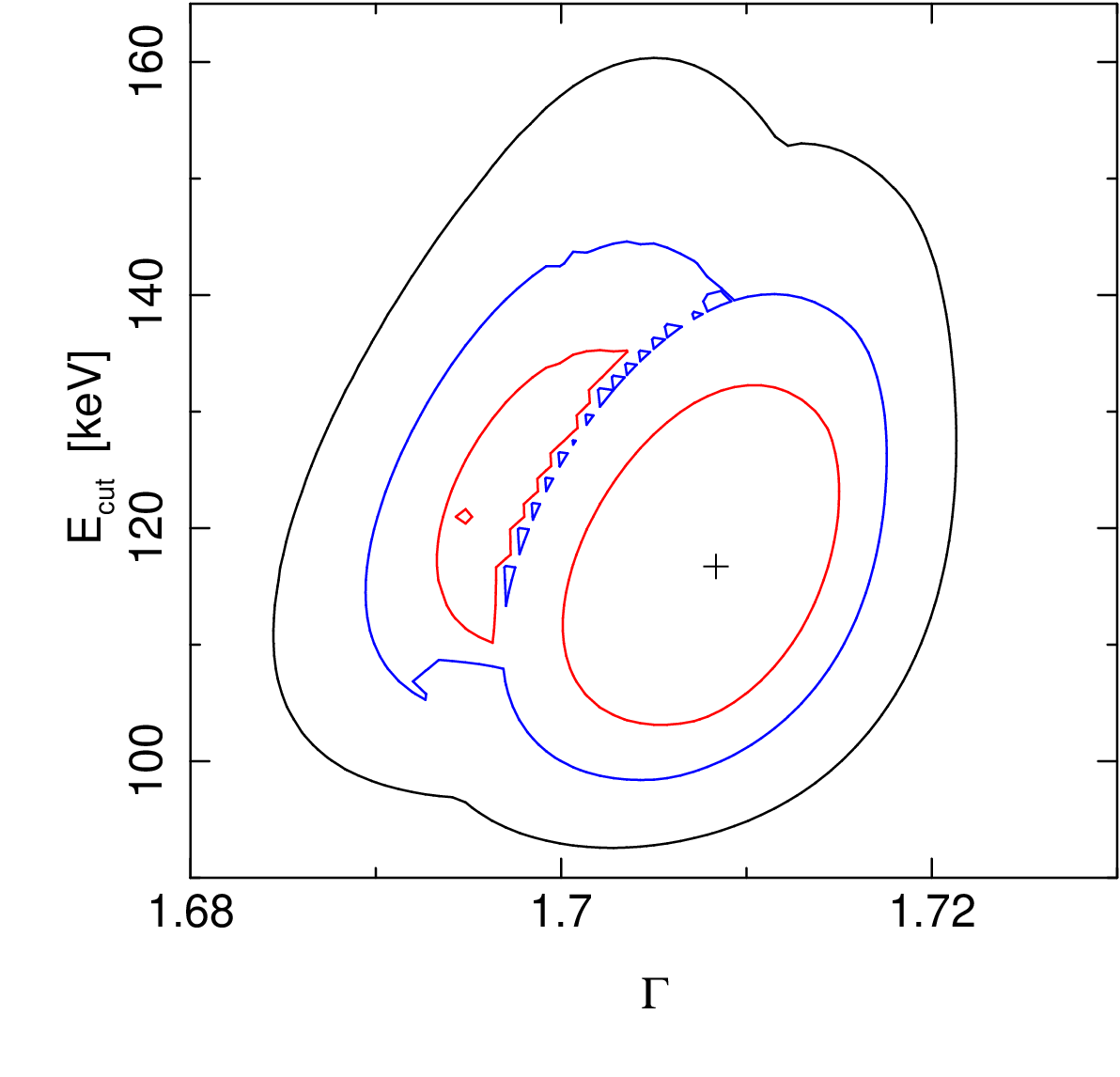}
\caption{Confidence contour of the photon index $\Gamma$ and the high energy cut-off energy $E_\text{cut}$.}\label{gamma_ecut}
\end{minipage}
\end{figure*}

\begin{figure*}[ht!]
%\vspace*{2.5\baselineskip}
\begin{center}
\includegraphics[width=2\columnwidth]{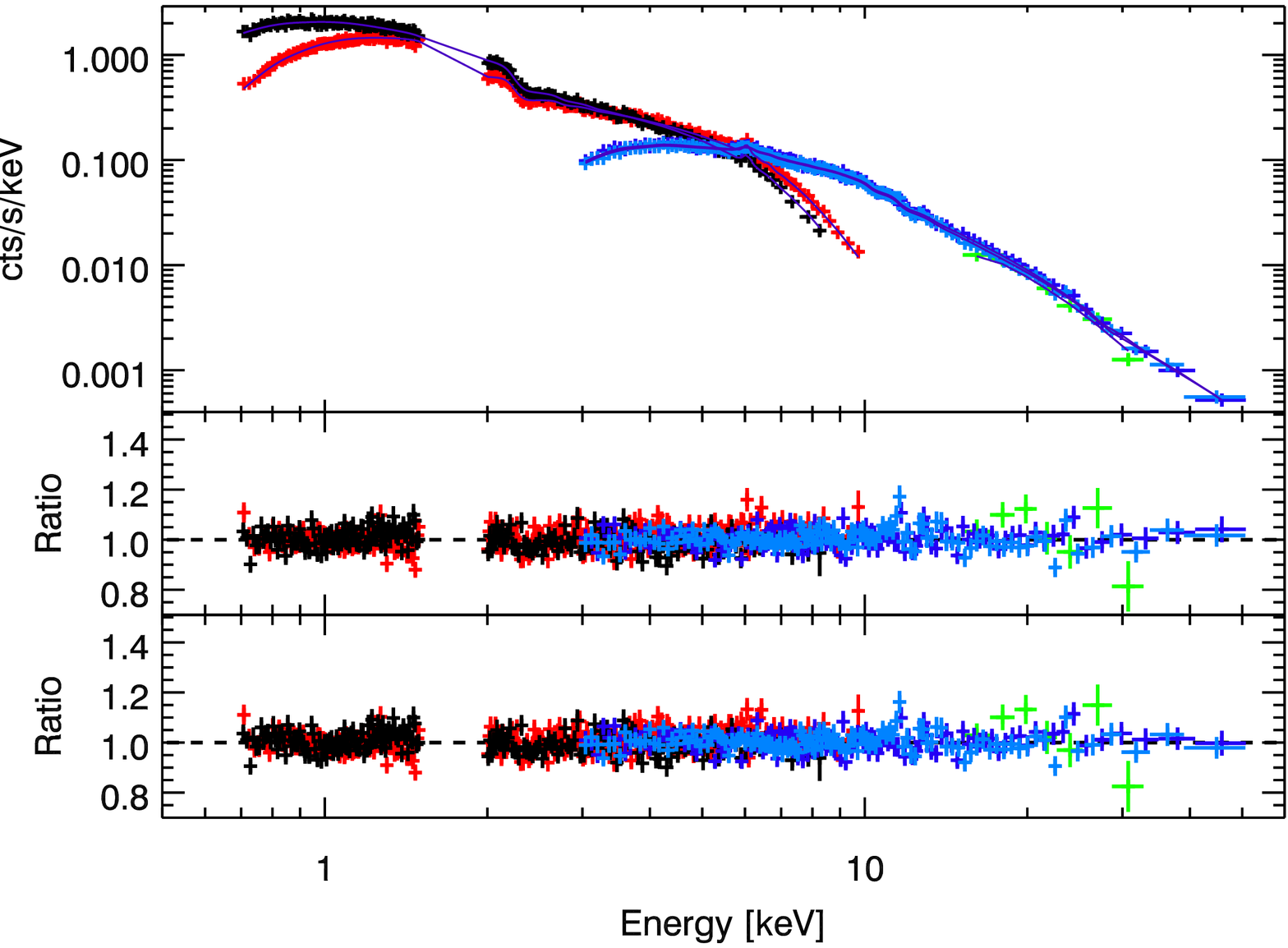}
\end{center}
\caption{The data [top panel] and residuals to the best-fitting \texttt{pexmon} [middle panel] and \texttt{eqpair} [bottom panel] model. In all cases the \textit{Suzaku}-XIS data are shown in black and red, the \textit{Suzaku}-PIN data in green, the \textit{NuSTAR} data in blue and the model in the top panel is shown in purple. The data have been rebinned in \texttt{xspec} for clarity.}\label{residuals}
\end{figure*}

The detailed, \textbf{final} results for this description of the data are shown in Table~\ref{3c390_fit} and residuals in Figure~\ref{residuals}. To ensure the robustness of our high energy cut-off constraints we compute contour plots with the neutral reflection fraction and photon index (Figures \ref{r_ecut} and \ref{gamma_ecut}). For the reflection fraction no degeneracy with the high energy cut-off is observed, highlighting the power of the \textit{NuSTAR} data to disentangle the primary continuum spectral shape from the contributions due to reflection. Equally the determined photon index is not degenerate with the high energy cut-off energy, however there seem to be two different solutions for the photon index indicating that the photon index is itself slightly degenerate with another parameter. 

Finally, we test whether the fit can be improved further by the inclusion of an additional power law component. Physically this is motivated by a possible jet contribution to the spectrum or a second coronal component that causes the weak soft excess. However, we find that no further improvement to the fit can be made this way. Replacing the ionized reflection altogether with just a power law also does not yield the same fit quality as ionized reflection provides.  

With already a good description of the data at hand we are now ready to replace the phenomenological power law continuum with a more physical continuum model. As the continuum in 3C\,390.3 is thought to be produced via the Comptonization of the soft accretion disk photons in a hot corona, we replace the power law with a thermal Comptonization model. Although, we are able to determine the high energy rollover, an indicator of the coronal temperature, with some accuracy the spectra are not able to distinguish different geometries for the corona. We therefore decide to model the Comptonization component with the \texttt{compTT} model \citep{Titarchuk1994} in the spherical coronal configuration. \texttt{compTT} describes the Comptonization spectrum with an analytic approximation assuming in our case a spherical corona where the seed photons (Wien distribution spectral shape) enter from the corona from the center. The Wien photon temperature was assumed to be 0.01\,keV\footnote{\textbf{As the obtained optical depth is rather high, i.e. a large number of scatterings takes place in the corona, the shape of the final Comptonization spectrum is not very sensitive to the exact seed photon temperature.}}. The \texttt{compTT} model does not include reflection and we therefore keep \texttt{pexmon} in our model, but freeze the photon index and norm to their previous best fit values. The power law that can be included via \texttt{pexmon} was also turned off, so that only the reflection spectrum remains. The \texttt{relxill} and \texttt{photemis} models are also included and operated as in the previous power law fit. Such a model yields a very good fit, we find the coronal temperature to be 16$_{-2}^{+4}$\,keV and the optical depth to be 5.5$_{-0.6}^{+0.4}$. The iron abundance has increased to 4.0 from 2.3 and the reflection fraction is slightly decreased. This can be understood as a trade-off between the curvature of the Comptonization and the reflection, the lack of curvature in the \texttt{pexmon} model led to overestimate of reflection. As the iron line strength remains the same, the abundance is higher in the Comptonization fit which includes overall less reflection. 

While \texttt{compTT} is a very popular thermal Comptonization model due to its simple parameterization and fast fitting speed, its accuracy is limited. We try to go beyond the simple analytic description of \texttt{compTT} and replace it with the more sophisticated model \texttt{eqpair} \citep{Coppi1999}. \texttt{eqpair} offers the advantage that it includes much of the physics of Comptonization without approximations and still each fit iteration is calculated quickly. Its drawbacks are that instead of treating the radiative transfer directly, it is summarized in a escape probability. \texttt{eqpair} cannot only treat thermal Comptonization but can also calculate hybrid and/or non-thermal Comptonization. For the purpose of this work, we limit ourselves to thermal Comptonization and switch the hybrid part of the model off. In contrast to \texttt{compTT} which is parameterized with respect to the electron temperature and optical depth, \texttt{eqpair} uses the compactness parameters. The hard and soft compactness parameter ratio $\ell_h/\ell_s$ plays the key role in determining the overall spectral shape, physically it indicates the amount of coronal heating to be balanced by the cooling provided by the seed photons entering the corona. The seed photon distributions available in \texttt{eqpair} are \texttt{diskbb} and \texttt{diskpn}. We have chosen \texttt{diskpn} with a peak temperature of 10\,eV for the modeling. The seed photons can be tweaked in the model by modifying the soft compactness parameter, however the data are not sensitive to this parameter so we fix it to 10. \texttt{eqpair} even incorporates the Compton hump from cold reflection, however it does not include the expected lines. We therefore shut off the reflection from \texttt{eqpair} itself by setting the reflection fraction to zero and include the reflection model components as we did with \texttt{compTT}. Similar to the fit with \texttt{compTT}, a fit with \texttt{eqpair} provides a good description of the data with a $\chi^2$ of 2564.3 for 2580 dof. Residuals for this fit are presented in the bottom panel of Figure~\ref{residuals}. We find a hard over soft compactness ($\ell_h/\ell_s$) of 69$_{-24}^{+124}$ and an optical depth of 4.1$_{-3.6}^{+0.5}$. The corresponding electron temperature is 30$_{-8}^{+32}$\,keV. 

\section{A note on spectral features in the iron K-band}\label{fe_modeling}

\begin{figure}[ht!]
%\vspace*{2.5\baselineskip}
\includegraphics[width=\columnwidth]{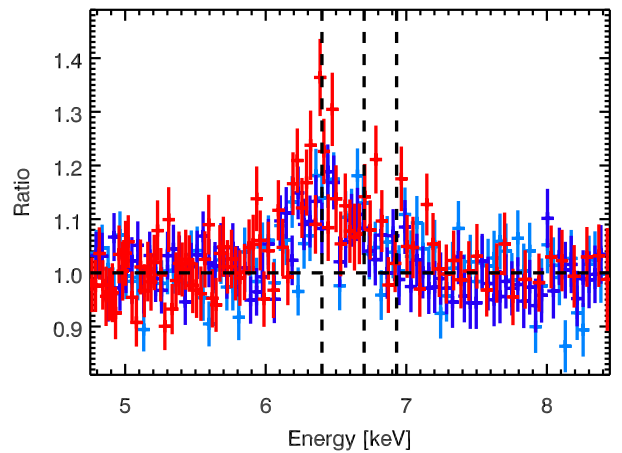}
\caption{\textit{Suzaku}-XIS [red] and \textit{NuSTAR} FPMA [light blue] and FPMB [blue] ratio to a simple power law fitted to the 4.5-5.5,keV and 6.5-8.0\,keV plotted in the rest frame of 3C\,390.3. Also indicated by the three vertical lines are the rest frame positions of Fe K$\alpha$, Fe XXV and Fe XXVI. \textit{NuSTAR} is binned to a signal-to-noise of 10 while \textit{Suzaku} is binned to a signal-to-noise of 17 to agree better with the resolution of \textit{NuSTAR}.}\label{iron_line_zoom}
\end{figure}

Having studied the broad-band spectrum, we now turn to the Fe K band in particular to search for additional structure that could either indicate the presence ionized reflection or possible blue-shifted absorption lines. To get a first impression of the detailed structure of the iron K region, we plot the \textit{Suzaku}/\textit{NuSTAR} residuals to a simple power law in 4.5 to 8.0\,keV energy band (Fig.~\ref{iron_line_zoom}). We observe a strong iron K$\alpha$ line and positive residuals close to the Fe XXVI rest frame energy.  

Our more detailed study of this region focuses on the 4-10~keV band of the observation-averaged combined \textit{Suzaku} front-illuminated XIS spectrum. For the continuum we assume a simple power law ($\Gamma$=1.68), as this describes the general spectral shape well in this region. We then test for the presence of emission and absorption lines using Gaussians. This way we confirm the presence of the neutral and highly ionized iron emission lines in the energy range $E\sim6.4-7$\,keV, already included in the broad band spectral modeling. We find the following properties for the emission lines: $E=6.37\pm0.02$\,keV and $\sigma=120\pm32$\,eV (EW$=58\pm6$\,eV) for the resolved Fe K$\alpha$ line and $E=6.86\pm0.06$\,keV and $\sigma<480$\,eV (EW$=24\pm6$\,eV) for the unresolved Fe XXV/XXVI complex, respectively. The equivalent widths were calculated with respect to the power law continuum. Considering the uncertainty on the line equivalent widths, we find that the lines are detected at $\simeq10\,\sigma$ and $\simeq4\,\sigma$ respectively, corresponding to fit improvements of $\Delta\chi^2/\Delta \text{dof}$ of 79/3, and 12/2. The final best-fit $\chi^2$/dof is 1169/1160. 

Additionally to the emission lines we see residuals in absorption at $E\simeq5$\,keV and $E\simeq7.5-8.5$\,keV in the \textit{Suzaku} spectrum, following the method of \citet{Tombesi2010}. However none of these residuals is significant at 3 or more $\sigma$ and the inclusion of the \textit{NuSTAR} spectra does not strengthen or confirm their detection.

The equivalent widths for the iron K$\alpha$ line observed here are in good agreement with the $56\pm16$\,eV found by \citet{Sambruna2009} for the earlier \textit{Suzaku} observation and the equivalent width of $68\pm14$\,eV determined by \citet{Tombesi2010} for the \textit{XMM} observation. \citet{Sambruna2009} also detect Fe XXV although it was observed to have a much larger equivalent width of about 80\,eV leading them to the conclusion that this was in fact a broad iron line.

\section{SED Modeling}\label{sed_modeling}

\begin{figure}[ht]
\includegraphics[width=\columnwidth]{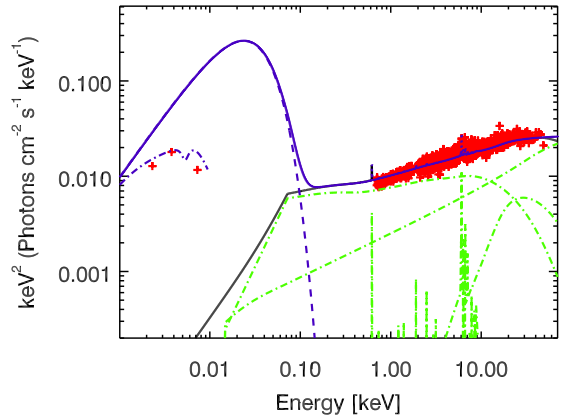}
\caption{Spectral Energy Distribution of 3C\,390.3 from \textit{Swift}-UVOT, \textit{Suzaku}-XIS, and \textit{NuSTAR}. The data are shown in red. The models overlayed are a) the X-ray Comptonization model shown in Table~\ref{3c390_fit} [black solid line], and b)the modified Comptonization model shown in Table~\ref{sed_modeling_table} [blue solid line]. The effects of Galactic absorption and reddening are not included in the plotted models, but the effect of intrinisic reddening is shown in the case of the modified Comptonization model as dotted-dashed blue curve. The dashed blue curve shows the multi-color disk blackbody unaffected by reddening. The other individual model components are shown in green.}\label{sed}
\end{figure}

So far, we have only focused on the X-ray data but a simultaneous \textit{Swift} pointing is also available, offering the opportunity to add the optical/UV data from UVOT. The Comptonization model \texttt{eqpair} used earlier does include the seed photon distribution, as well as any unscattered fraction of seed photons. Therefore, assuming the accretion disk photons are upscattered in the hot corona, one would expect an extrapolation of the best fit spectral model found from the X-ray modeling to describe the UV/optical as well. We find that the UV/optical flux predicted from the seed photons, does underpredict the flux in the UV/optical by several orders of magnitude (Figure~\ref{sed} black curve). Not only does the X-ray best fit model not predict enough UV flux but the shape of the model does also not resemble the data very well. This could possibly be due to intrinsic reddening in the AGN host galaxy. which has not yet been accounted for. Based on these observations two modifications are made to the spectral model: 1) an additional multi-color disk black body is in the modeling to account for the flux difference (its temperature is fixed to that of the seed photons), and 2) the \texttt{zdust} model is included to account for intrinsic reddening. The method in the \texttt{zdust} model is set to 2 (LMC) and the $R(V)$ value to 3.16. We adopt an E(B-V) value of 0.205, which is the best fit value from the Balmer decrement by \citet{Dietrich2012}. The two modifications allow a good description of the data considering that the UV/optical photometry values have very small and possibly underestimated errors. The changes to the Comptonization parameters themselves are marginal see Table~\ref{sed_modeling_table}. The additional disk has a norm of 3.9$_{-0.1}^{+0.1} \times 10^9$.   

\begin{table}[h]
\caption{Spectral fit parameters from a fit with a Comptonization model, including also cold reflection and an accretion disk, to \textit{Swift}, \textit{Suzaku} and \textit{NuSTAR} data of 3C\,390.3. Parameters marked with an 'f' have been kept fixed at their previous best fit values.}\label{sed_modeling_table}
\begin{center}
\begin{tabular}{c|c|c}
\hline \hline Cold Reflection & $\Gamma$ &  1.71f \\
 & $E_\text{cut}$ [keV] &  117f \\
 & $R$ &  0.31$_{-0.03}^{+0.04}$\\
 & Fe/Solar &  0.8$_{-0.1}^{0.9}$ \\
 & $N_\text{pex}\,[10^{-2}]$ &  1.00f \\
\hline Comptonization & $kT_e$ [keV] &  42$_{-18}^{+20}$ \\ 
 & $\ell_h/\ell_s$ & 85$_{-27}^{+111}$ \\
 & $\tau$ & 3.3$_{-2.8}^{+1.3}$  \\
 & $N_\text{eqpair}$ & 94$_{-40}^{+53}$ \\
\hline Ionized Reflection & $a$ & 0.993$_{-0.003}^{+0.004}$ \\
& $q$ & $>9.8$ \\
& $\log(\xi)$ & 3.52$_{-0.06}^{+0.12}$ \\
\hline Photoemission & $N_\text{phot}$ & $>0.002$ \\
                & $\log(\xi)$ & 3.9$_{-0.1}^{+0.3}$ \\ 
\hline Intrinsic Reddening & $E(B-V)$ & 0.205f \\
\hline Accretion Disk & $N_\text{disk}$ [$10^9$] & 3.9$_{-0.1}^{+0.1}$ \\
\hline Cross Calibration & $c_\text{xis1}$ & 0.977f\\
  & $c_\text{pin}$ & 1.16f\\
  & $c_\text{fa}$ & 1.00f \\
  & $c_\text{fb}$ & 1.05f\\
\hline \hline  & $\chi^2$ & 2709.4 \\
  & dof & 2582 \\
\end{tabular}
\end{center}
\end{table}

We also tested an alternative scenario similar to the one suggested by \citet{Petrucci2013}, where the accretion disk is covered by a warm layer producing a low temperature, optically-thick Comptonization component that extends from the optical/UV into the soft X-ray band. This component is assumed to replace the standard accretion disk component. However, we find that such a component is unable to produce enough optical/UV flux without producing significantly more soft X-ray flux than observed.

\section{Discussion}\label{disc}

With a 2-10\,keV flux of $4.03_{-0.01}^{+0.02}\times10^{-11}\,\text{ergs}\,\text{s}^{-1}\,\text{cm}^{-2}$ the source was observed to be brighter in this observation than in any other previous high quality CCD data and slightly above the average flux of all previous \textit{RXTE} observations \citep{rivers2013}. From the 2-10\,keV flux, we estimate an Eddington fraction of 1\,\% using a bolometric correction of 6.33 as determined by \citet{Vasudevan2009}. The results from the X-ray spectral analysis in this work are in good agreement with those of the previous broad-band analysis by \citet{Sambruna2009}. Contrary to the softer-when-brighter trend seen in many Seyferts, the photon index of the power law in this observation is similar to that of the previous high quality CCD observations at lower flux ($\Gamma=1.71$ versus 1.72), suggesting a rather stable shape for the primary continuum. The neutral reflection fraction on the other hand is found to be much smaller in this observation ($R=0.14$ versus 0.81) and much more in the realm that is typically expected for broad-line radio galaxies. This could however be due to the inclusion of the BAT data, which is non-simultaneaous in the \citet{Sambruna2009} analysis. Attributing the discrepancy to such a technical problem seems reasonable given that the average PCA+HXTE spectra indicate reflection fraction of $R=0.2\pm0.1$ \citep{rivers2013}. In our observation we do not find evidence for any broadening in the iron K$\alpha$ line (\S\ref{fe_modeling}), however we do find a weak soft excess that can be well described by a blurred ionized reflector. Attempting to model the soft X-ray excess with an additional power law, which could physically represent either a low energy Comptonization component \citep{Done2012,Petrucci2013} or a jet component, leads to a worse fit than the reflection scenario. (This is also confirmed by the SED fitting in \S\ref{sed_modeling}.) Therefore we conclude similar to \citet{Sambruna2009} and \citet{Walton2012} that an ionized reflector is present in the X-ray spectrum of this AGN. The fact that it is not very apparent in this observation can be explained by a variable ionized reflector. A variable ionized reflector is predicted along with the formation of the new jet knots \citep{Marscher2002,Chatterjee2009,Lohfink2013} and would therefore not be surprising. The exact strength and shape of the ionized reflection component in 3C\,390.3 depends on the Comptonization model, as we will see below. The overall exceptionally low reflection fraction observed in broad-line radio galaxies can be explained by an outflowing corona \citep{Beloborodov1999,Malzac2001}. If the corona is outflowing at mildly relativistic speeds, the emission is beamed away from the accretion disk lowering the amount of reflection.

\begin{figure}[ht]
\includegraphics[width=\columnwidth]{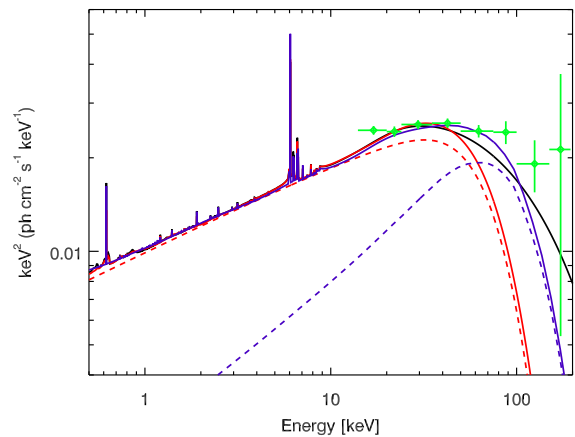}
\caption{70 month \textit{Swift}-BAT spectrum with overlayed unabsorbed best fit models for the three different continua considered here: \texttt{pexmon} (black curve), \texttt{compTT} (red curve), \texttt{eqpair} (purple curve). The dashed curves show only the primary continuum for the two Comptonization fits.}\label{models_comparison}
\end{figure}

Overall, we find that the X-ray spectrum of 3C\,390.3 can be well described by a model consisting of a primary continuum, cold and ionized reflection, and photoionized emission. For the primary continuum we consider three possibilities: a cut-off power law as a purely phenomenological model and the two thermal Comptonization models \texttt{compTT} and \texttt{eqpair}. All continua yield fits of comparable quality and the total spectral models consequently appear almost identical in the observable 0.7-80\,keV band (Fig.~\ref{models_comparison}). However, the contribution of the different model components differs for the two Comptonization models. In Figure~\ref{models_comparison} the dashed lines show the primary continuum for both cases, indicating that for the \texttt{compTT} case the primary continuum accounts for most of the flux at lower energies, while for \texttt{eqpair} this is not the case. The remaining, required soft X-ray flux in the \texttt{eqpair} case is made by neutral and ionized reflection, as the increase in  the reflection fraction shows with $R_\text{neu,comptt}$=0.09 to $R_\text{neu,eqpair}$=0.30 and $R_\text{ion,comptt}$=0.07 to $R_\text{ion,eqpair}$=0.23. At first glance one might therefore be tempted to reject the \texttt{eqpair} solution. The view changes, however, as one extends the models to energies above 80\,keV, where the models diverge. No X-ray data exists for those energies that are simultaneous with the \textit{Suzaku}/\textit{NuSTAR} observation, but we can use the average 70\,month \textit{Swift}-BAT spectrum as guidance on how the X-ray spectrum might look like at higher energies. The spectrum, which is overlayed on the best fit model in Figure~\ref{models_comparison}, agrees well with all best fit models below 40\,keV, above this energy it is systematically brighter. It is important to note here that it is inherent to Comptonization models that they roll over more rapidly than the phenomenological exponential cut-off included in \texttt{pexmon} \citep{Zdziarski2003}. Comparing the two Comptonization models, it seems that the \texttt{eqpair} solution produces fluxes better in agreement with BAT spectrum. 

The high energy cut-off of our thermal Comptonization spectrum of 117$_{-14}^{+18}$\,keV is in very good agreement with a joint IBIS/BAT spectral analysis suggesting a cut-off temperature of $97^{+20}_{-11}$\,keV \citep{Malizia2014} and the result from the joint EPIC-pn/XIS/PIN/BAT fit by \citet{Sambruna2009} returning $E_\text{cut}=157^{+89}_{-47}$\,keV. This would suggest that the coronal temperature was constant, however the large errorbars cannot exclude a change. Possibly, Fig.~\ref{models_comparison} could be seen as weak evidence that at brighter fluxes the cut-off is lower. Such a trend could be caused by increased coronal cooling from stronger UV emission (more seed photons). Indeed, when comparing the UVOT fluxes to the OM fluxes we find that the source was brighter in the new UVOT observation (the AGN flux doubled in the U-band). It is likely that given the large errorbars a small change in the coronal temperature/high energy cut-off would be undetectable. 

The coronal temperature found for 3C\,390.3 is rather cool ($30_{-8}^{+32}$\,keV), similar to what has been found in IC 4329A \citep{Brenneman2014}, SWIFT J2127.4+5654 \citep{Marinucci2014} and MCG-05-23-016 \citep{Balokovic2015}. In contrast to these coronae with temperatures of $<100\,\text{keV}$, \textit{NuSTAR} has also found several sources with coronal temperatures above 150\,keV such as 3C\,382 \citep{Ballantyne2014} and Ark\,120 \citep{Matt2014}. The number of measurements is still too small to suggest a reason for the wide spread of the measured temperatures. One possibility could be that the coronal temperature is linked to the radiative compactness of the corona. While we have already noted that the hard over soft compactness measures the ratio of heating and cooling, the compactness by itself tells us about the importance of photon-photon interactions \citep{Guilbert1983}. If $\ell$ is much greater than 1 (as is the case here) these become important. If sufficient high-energy photons are produced via Compton upscattering these will start to create electron/positron pairs until at some point there will be so many pairs formed that the temperature of the corona is decreased significantly. Therefore there is a maximum temperature a corona can reach at any given compactness. Below this temperature the corona would still be able to heat up further, but above it pair production will cool it down again. Accurate predictions of this equilibrium temperature are difficult as it depends on the unknown geometrical layout of the disk corona system. However, \citet{Fabian2015} show that the coronal temperature measurements with \textit{NuSTAR} do indeed show a dependency with the coronal compactness that is similar to that expected for a slab corona. The measurement of 3C\,390.3, which is also included in \citet{Fabian2015}, agrees well with the other measurements and supports the idea of an equilibrium temperature.

Aside from the coronal temperature the other key parameter returned by the Comptonization models is the optical depth of the corona. From our measurements we find it to be large ($\sim 3$), meaning that all soft photons entering the corona will be scattered multiple times. Depending on the geometry of the corona this could lead to an apparent truncation of the accretion disk as the photons cannot escape unscattered. This apparent truncation could be  detected in the reflection spectrum and the accretion disk spectrum. Unfortunately the \textit{Swift}-UVOT photometry does not allow a measurement of the accretion disk temperature for the unscattered component, which could indentify a blocked central region. However, our measurements of high spin and very high emissivity from the reflection spectrum, if taken at face value, suggest that we are seeing very close to the black hole. In fact the vast majority of the reflection would stem from a small ring around the black hole. The easiest way to bring this into agreement with the high optical depth of the corona would be a very compact corona (see also below). We caution however that the extreme reflection parameters could also indicate that we have reached the limits of the current spectral models.

Curious is also the disagreement between UV predicted by the X-ray spectral and measured UV emission by about a factor of $\sim20$. A similar phenomenon has been found for 3C\,382 with also a factor of about 10 \citep{Ballantyne2014}. This mismatch of predicted and measured UV/optical emission can be explained by the spatial compactness of the corona, which causes only a small fraction of the total disk emission to enter the corona and be Comptonized. There is much observational evidence for the corona being of small spatial extent, for example microlensing \citep[e.g.,][]{Dai2004,Chartas2009}, broad iron lines \citep[e.g.,][]{Fabian2013,Parker2014}, and time lags \citep[e.g.,][]{Zoghbi2010,Kara2013}. The SED presented in Figure~\ref{sed} also reveals that a naive modeling of the SED as is performed in this work predicts detectable accretion disk flux in the very soft X-ray band. This small flux contribution from the accretion disk is inconsistent with the weak soft excess observed in the X-ray spectrum of 3C\,390.3. %For a black hole of the mass of 3C390.3 this is unphysical, however without a better high quality UV spectrum it is hard to determine whether the simplicity of disk model or another issue is causing this problem. 

With the continuum being describable by only thermal Comptonization, no contribution from the jet to the X-ray spectrum can be found in 3C\,390.3. This is another matter concerning radio galaxy X-ray spectra, which is generally still open to debate \citep[e.g.,][]{Beckmann2011}. Our finding is in good agreement with the Fermi non-detection of 3C\,390.3 suggesting that the jet emission is only very weak in the hard X-ray band \citep{Kataoka2011}.

\acknowledgments
\section*{Acknowledgments}
We thank the anonymous referee for their helpful comments. AL thanks Julien Malzac for helpful discussions and acknowledges support from the ERC Advanced Grant FEEDBACK. FT would like to thank M. Coleman Miller and Brian Morsony for the useful comments. M.\,B. acknowledges support from NASA Headquarters under the NASA Earth and Space Science Fellowship Program, grant NNX14AQ07H. This work made use of data from the \textit{NuSTAR} mission, a project led by the California Institute of Technology, managed by the Jet Propulsion Laboratory, and funded by the National Aeronautics and Space Administration. This research has made use of the \textit{NuSTAR} Data Analysis Software (NuSTARDAS) jointly developed by the ASI Science Data Center (ASDC, Italy) and the California Institute of Technology (USA). This research has made use of data obtained from the \textit{Suzaku} satellite, a collaborative mission between the space agencies of Japan (JAXA) and the USA (NASA).

\bibliographystyle{apj}

\end{document}